\documentclass[12pt]{article}
\usepackage{epsf}
\def\be{\begin{equation}}
\def\ee{\end{equation}}
\def\bea{\begin{eqnarray}}
\def\eea{\end{eqnarray}}

\begin{document}
\begin{titlepage}
\begin{center}
{\Large \bf William I. Fine Theoretical Physics Institute \\
University of Minnesota \\}
\end{center}
\vspace{0.2in}
\begin{flushright}
FTPI-MINN-15/34 \\
UMN-TH-3443/15 \\
July 2015 \\
\end{flushright}
\vspace{0.3in}
\begin{center}
{\Large \bf Modification by pion exchange of near threshold resonance line shape in open heavy flavor channel. 
\\}
\vspace{0.2in}
{\bf M.B. Voloshin  \\ }
William I. Fine Theoretical Physics Institute, University of
Minnesota,\\ Minneapolis, MN 55455, USA \\
School of Physics and Astronomy, University of Minnesota, Minneapolis, MN 55455, USA \\ and \\
Institute of Theoretical and Experimental Physics, Moscow, 117218, Russia
\\[0.2in]

\end{center}

\vspace{0.2in}

\begin{abstract}
The effect of the pion exchange on the line shape of near threshold bottomoniumlike resonances decaying into an $S$ wave pair of $B^{(*)}$ mesons is considered. It is pointed out that this effect, parametrically enhanced by the heavy meson mass,  can be of a practical significance in determining the parameters of the bottomoniumlike resonances such as the known $Z_b(10610)$ and $Z_b(10650)$ states. 
\end{abstract}
\end{titlepage}

Experimental studies of the charm and bottom threshold regions have uncovered a number of quarkoniumlike peaks with mass very near a threshold of a heavy meson-antimeson pair. In particular, the bottomoniumlike $Z_b(10610)$ and $Z_b(10650)$~\cite{bellez} peaks are located at the respective thresholds, $B B^*$ and $B^* B^*$, and within the present accuracy of a few MeV it is impossible to reliably conclude whether the mass of each of this peaks is below, at, or above the corresponding threshold. These structures, being isotopic vectors, definitely contain a light quark-antiquark pair in addition to the heavy pair $b \bar b$. Their properties agree well~\cite{bgmmv} with the molecular~\cite{ov} picture where the $Z_b$ resonances are considered as shallow bound states of the corresponding meson-antimeson pair, and also alternative models for these states are being discussed in literature, e.g. a tetraquark scheme~\cite{mppr}  has been used~\cite{ahw,ampr} for description of hidden-beauty states, including the $Z_b$ resonances. Other bottomoniumlike resonances of similar nature are expected at meson-antimeson thresholds. In particular, in the molecular scheme two or four additional isovector states are expected~\cite{mv11} that were not yet observed due to their opposite, relative to $Z_b$, $G$-parity. In either model of the internal four-quark dynamics within such resonances,  these peaks present a major feature, a bound state or a virtual state near-threshold pole, in the $S$ wave scattering of a heavy meson and antimeson. For the $Z_b$ resonances a strong coupling between the $Z_b$ states and the meson pairs has been in fact observed experimentally~\cite{bellezbb} with the spectral density in the  $B^* \bar B + B \bar B^*$ channel displaying a peak corresponding to the $Z_b(10610)$ and that in the $B^* \bar B^*$ channel has a peak corresponding to the $Z_b(10650)$. Clearly, a study of the line shape in these open-flavor channels as well as in the decay channels with bottomonium and light meson(s) is necessary for determining the masses and widths of the resonances precisely enough for clarifying the type of the features that these resonances correspond to in the meson-antimeson dynamics and, possibly, for understanding the internal structure of these states. This however requires also a refined theoretical description of the relation between the observed line shapes and the parameters of the interaction between the heavy mesons. 

Most recently a practical improvement on a straightforward summation of Breit-Wigner amplitudes has been suggested~\cite{hkmmnw},  which accounts for the coupling between channels and its effect on the interference of the $Z_b(10610)$ and $Z_b(10650)$ resonances. The energy scale for the modification of the line shapes of the resonances considered in Ref.~\cite{hkmmnw} is set by the gap between the thresholds $\Delta=M(B^*)-M(B) \approx 46\,$MeV, or equivalently by the mass difference between the two $Z_b$ resonances. The present paper addresses a different type of modification of the line shape within each individual peak which modification significantly changes at an energy scale $\delta \sim \mu^2/M$ set by the pion mass $\mu$ and the meson mass $M$, and is induced by the interaction between the mesons through pion exchange. The latter correction to the yield of the meson pairs produced with the center of mass (c.m.) momentum $p$ is proportional to the ratio $M/p$ if $p$ is larger than $\mu$ (this is similar to the $1/v$ behavior of the well known Coulomb correction to the yield of charged particles) and this ratio is replaced by $M/\mu$ at $p$ being smaller than $\mu$. The discussed effect is determined by the interaction at characteristic distances of order $1/p$ or $1/\mu$ respectively, and thus is sensitive to only the pion exchange. Any effects of the interaction at shorter distances, of order $1/\Lambda_{QCD}$ display variation at a larger energy scale $ \Lambda_{QCD}^2/M \sim \Delta$ and can be coded in terms of the `short distance' scattering parameters for the mesons, similarly to the approach described in Ref.~\cite{hkmmnw}. The separation of the `long distance' pion-mediated force and the short distance effects is thus possible as long as $p^2 \ll \Lambda_{QCD}^2$, i.e. in the range of excitation energy up to approximately 10\,MeV for the $B$ meson pairs~\cite{mv13,lv}. Numerically, the effect of the pion exchange gives a relative correction of approximately 30\% to the yield of $S$-wave pairs in the isovector channel at the very threshold and rapidly falls off at the excitation energy of order $\delta$. As will be seen from what follows, the detailed behavior of the discussed correction also depends on the elastic scattering amplitude for the $B$ mesons generated by the interaction at short distances of order $1/\Lambda_{QCD}$. Therefore a study of the behavior of the yield can provide an insight into the properties of this interaction. The calculated modification of the production of $S$-wave pairs of heavy mesons is also applicable in the case where there is no resonance at the threshold. However it appears to be most interesting in the channels where such resonance exists, and can be of significance for studying the properties of the resonance peak.  In practice the effect of a relatively rapid variation of the correction to the yield can be measured from the ratio of the production of the threshold meson pairs and one of non-threshold inelastic channels, such as e.g. the ratio of the final states $B^* \bar B + B \bar B^*$ and $h_b \pi$ at the $Z_b(10610)$ resonance, or the ratio of $B^* \bar B^*$ to $B^* \bar B + B \bar B^*$ at the $Z_b(10650)$. The non-threshold channels are not influenced by the rapidly varying with energy correction, and the overall variation of the yield due to the underlying resonance cancels in such ratios.

The interaction between slow heavy mesons can be viewed as consisting of two parts: the short distance part and the long distance one due to pion exchange. The short distance part gives rise to both the elastic scattering and to inelastic processes, i.e. this part of the interaction is responsible for the existence of the near-threshold resonances and also for their decays into `other' channels, such as into bottomonium plus light mesons, or annihilation into light hadrons. It can be noted that the mixing between channels with different heavy meson pairs, e.g. between $B^* \bar B^*$ and $B^* \bar B + B \bar B^*$ should be treated as an inelastic process induced by a short distance interaction, since the c.m. momentum (real or imaginary) for one pair at the threshold of the other is $p_0 \approx \sqrt{\Delta M} \approx 0.5\,$GeV and is thus of order $\Lambda_{QCD}$, both numerically and parametrically. In the terminology used in the recent literature the short distance interaction is described by contact terms in the effective Lagrangian for the heavy mesons~\cite{mp} (see also Ref.~\cite{hkmmnw} and references therein). An equivalent standard approach to treatment of a  strong interaction, pursued in textbooks (e.g. in Ref.~\cite{ll}), introduces scattering parameters generated by the short range forces. For the present calculation the relevant parameters are the diagonal (elastic) element $S_e$ of the $S$ matrix and the effective range $a$ of the short distance interaction. The latter parameter separates the short and the long distances: at $r < a$ there is strong interaction generating the scattering matrix, while at $r > a$ there is only the long range force, described for the purposes of the present calculation by the pion exchange. The modification of the scattering by the interaction at long distances is  then found by matching at $r=a$ the `outer' wave function to the scattering state wave function described by the $S$ matrix. Clearly, this picture of an `abrupt' separation between the short and long distances is an approximation, and the parameter $a$ should be treated as an effective one~\footnote{In the effective Lagrangian approach the corresponding quantities are the form factors for the contact terms.}. Moreover the whole approach is applicable as long as the dependence of the final result on precise definition of $a$ is not critical. 

The long range forces induced by the pion exchange are determined from the Lagrangian for the interaction of the vector ($V$) $B^*$ mesons and the pseudoscalar ($P$) $B$ mesons, as isotopic doublets, with the pion isotopic triplet $\pi^a$:
\be
H_{int} = {g \over f_\pi} \, \left \{ \left [ (V^\dagger_l \tau^a P) + {\rm h.c.} \right ]+ i \, \epsilon_{ljk} \, (V^\dagger_j \tau^a V_k) \right \}  \,  \partial_l \pi^a
\label{lint}
\ee
with $\tau^a$ being the isospin Pauli matrices, and the nonrelativistic normalization is implied for the heavy mesons. The charged pion decay constant $f_\pi \approx 132\,$MeV is used in Eq.(\ref{lint}) for normalization. The dimensionless pion coupling $g$ can then be evaluated by using the heavy quark symmetry and the known~\cite{cleo} (and recently updated~\cite{babar,pdg}) absolute rate of the $D^{*+} \to D \pi$ decay: $g^2 \approx 0.15$. 

The forces due to the pion exchange generally give rise to several effects~\cite{nv}: a central interaction potential, a tensor potential, and a mixing between channels related by interchange of the vector and pseudoscalar mesons. 
The effect of the pion exchange is considered here for an $S$-wave state of a heavy meson pair as a perturbation in the first order (an approximation to be justified by the result). For this reason only the long range part of the central potential is relevant for the present discussion. This part is given by~\cite{nv}
\be
V(r) =  c \, \left[2 I (I+1) - 3 \right ] \, {g^2 \over 12 \pi}  \, {\mu^2 \over f_\pi^2} \, {e^{- \mu r} \over r}~,
\label{vc}
\ee
where $I$ is the isospin of the meson pair and the constant $c$ is nonzero only if at least one of the mesons is a vector (i.e. naturally $c=0$ in a $B \bar B$ system) and depends on the total spin $J$ of the pair and its symmetry under the charge conjugation: $c= 1$,  $2$, $-1$ and $-1$ respectively for $J^{PC} = 1^{+-}$, $0^{++}$, $1^{++}$ and $2^{++}$, with $C$ standing for the charge parity of the neutral component of the isotopic multiplet. In particular, for the case of the $Z_b$ resonances ($I=1$, $J^{PC} = 1^{+-}$) this potential is a repulsion given by
\be
V_Z = {g^2 \over 12 \pi}  \, {\mu^2 \over f_\pi^2} \, {e^{- \mu r} \over r}~.
\label{vz}
\ee
It should be also noted that in the systems made of a vector and a pseudoscalar heavy mesons the parameter $\mu$ is not exactly the pion mass $m_\pi$ but rather is $\mu = \sqrt{m_\pi^2 - \Delta^2} \approx 127\,$MeV (for the exchange of $\pi^0$, which is the case for the $Z_b^\pm$ systems) due to the transfer of the energy $\Delta$ between the mesons in the $B^* \bar B \to B \bar B^*$ rescattering~\cite{nv11,mpv,nv}.

The modification by a long range interaction of the yield of a final state generated by a source localized at shot distances $r < a$ has been considered in a general form in Ref.~\cite{dlorv} within the described above setting for separation between the distance scales (see also in Ref.\cite{mv12}). The rate $\Gamma_0$ that would give the yield in the absence of a long range force is replaced by $\Gamma=R \, \Gamma_0$ with the correction factor $R$ given by
\be
R= 1- { M \over p} \, {\rm Im}\left [S_e(p) \,
\int_a^\infty e^{2ipr}  \,
V(r) \, dr \right ]~,
\label{rf}
\ee
where $S_e$ is the elastic element of the $S$ matrix in the considered channel, $p$ is the c.m. momentum for each component of the heavy pair, and $M$ is twice the reduced mass (i.e. $M$ is approximately the mass of a heavy meson). 

A substitution of the potential from Eq.(\ref{vc}) in the formula (\ref{rf}) for the correction factor readily produces the result for the discussed modification of the yield by the pion exchange. In order to estimate the magnitude of the effect it is instructive to consider some limiting cases. The correction is the largest in the limit $p \to 0$~\footnote{At this point, i.e. exactly at the threshold, the phase space for the pair also goes to zero linearly in $p$, so that the consideration in this limit has to be applied to the slope of the yield of the pairs.} where there is no uncertainty due to the scattering matrix element, since $S_e(0) = 1$ (see e.g. in the textbook \cite{ll}), so that one estimates
\be 
R = 1 - c \, \left[2 I (I+1) - 3 \right ] \, {g^2 \over 12 \pi}  \, {\mu^2 \over f_\pi^2} \, {2 M \over \mu} \, e^{-\mu a} \approx 1- 0.3 \, c \, \left[2 I (I+1) - 3 \right ] ~.
\label{zerop}
\ee 
The effective radius of the strong interaction $a$ is formally considered small in comparison with $1/\mu$, so that the deviation of the exponent from one is beyond the accuracy of the discussed approach. One can readily see from this estimate that for the $I=1$ states with the quantum numbers $1^{+-}$, $1^{++}$ and $2^{++}$ the correction has a moderate value amounting to about 30\% exactly at the threshold, which justifies the treatment of the pion exchange at long distances as a perturbation. This includes the currently most interesting case of the $Z_b$ resonances, for which the correction at the threshold is negative. Clearly, the estimate in Eq.(\ref{zerop}) gives for the $2^{++}$ $I=1$ state and for all the states with $I=0$ the value of the correction that is too large to justify its treatment in the first order of perturbation theory. In these cases the treatment of the effect of the pion exchange at energies all the way down to the threshold should definitely be refined. However, the linear in $V$ expression can still be applied at $p$ somewhat above the threshold where the correction is smaller.  It can be also noted  that the isoscalar states are likely to mix with pure $b \bar b$ bottomonium, so that it would be a coincidence if an isoscalar bottomoniumlike resonance was found within a few MeV from the threshold. Another remark, related to the formula in Eq.(\ref{zerop}) is that for the charmed mesons the effect is smaller by the factor $M(D)/M(B)$ and appears to be of less phenomenological significance than for the $B$ mesons.

Proceeding to the case of a finite momentum $p$,  the leading at small $a$ behavior of the correction can be found in terms of the real and imaginary parts of $S_e$ as
\be
R = 1 - c \, \left[2 I (I+1) - 3 \right ] \, {g^2 \over 12 \pi}  \, {\mu^2 \over f_\pi^2} \, {M \over p} \, \left [ {\rm Re} S_e \, \arctan {2p \over \mu} - {\rm Im} S_e \, \log \left ( a \, \sqrt{\mu^2 + 4 p^2} \right ) \right ] ~.
\label{rfull}
\ee
This expression implies that a non removable weak logarithmic dependence on the parameter $a$ arises as soon as Im$S_e$ is nonzero, which dependence is well known in similar problems with $S$ wave scattering~\cite{ll}~\footnote{One can readily notice that for $S_e=1$ and in the limit $\mu \to 0$ the formula in Eq.(\ref{rfull}) reproduces the familiar $1/v$ behavior for the effect of the Coulomb interaction. It can be also noted that no singularity arises at $p=0$ from the term with Im$S_e$, since Im$S_e \propto p$ at $p \to 0$.}. 

The low-momentum behavior of the elastic scattering amplitude can be parametrized as  $f_e = -1/(\kappa + i \, p)$ with $\kappa$ being the inverse of the scattering length, so that
\be
S_e = 1 + 2 i \,  p \, f_e = {\kappa - i \, p \over \kappa + i \, p}~.
\label{sl}
\ee
In the presence of absorption into inelastic channels, as is the case with the realistic bottomoniumlike resonances, the parameter $\kappa$ is generally complex: $\kappa = \kappa_0 + i \, \kappa_1$. A shallow bound state corresponds to a small positive $\kappa_0$ and the resonance peak is located at energy $\kappa_0^2/M$ below the threshold. A small negative $\kappa_0$ corresponds to a virtual state, and the peak of the spectral density is located exactly at the threshold. Currently it is not known which particular case applies to the observed $Z_b$ peaks. (Hopefully, a study along the lines discussed here can help in removing this uncertainty.)  Here I simply assume, for illustrative purposes, that the real part, $\kappa_0$, is small as compared to the imaginary, part $\kappa_1$. In this case the scattering parameter $S_e$ is then purely real: $S_e =(\kappa_1 - p)/(\kappa_1+p)$, and  $\kappa_1$ can be estimated from the measured width $\Gamma_Z \approx 15\,$MeV of the $Z_b$ resonances: $\kappa_1 = \sqrt{\Gamma_Z M/2} \approx 200\,$MeV. The resulting behavior of the correction to the yield of $B^* \bar B + B \bar B^*$ pairs near the threshold at $E_0=M(B^*) + M(B) \approx 10605\,$MeV is shown in Fig.~1. 

\begin{figure}[ht]
\begin{center}
 \leavevmode
    \epsfxsize=13cm
    \epsfbox{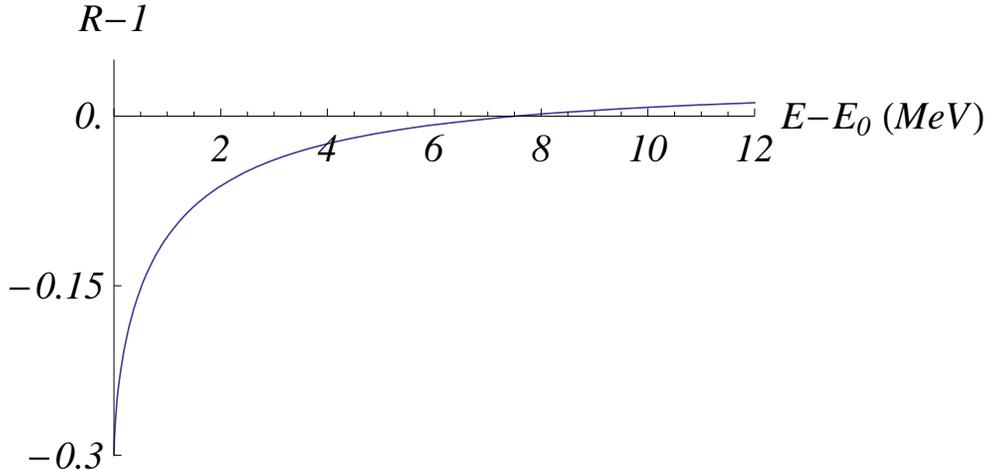}
    \caption{An illustrative plot of the relative correction to the yield of the $B^* \bar B + B \bar B^*$ pairs as a function of the c.m. energy $E$ above the threshold in the channel with the quantum numbers of the $Z_b(10610)$ resonance.}
\end{center}
\end{figure} 

In summary. The long range force generated by the pion exchange modifies  the excitation of $S$ wave pairs of heavy mesons in the immediate vicinity of their thresholds. The resulting correction to the yield, given by the first-order expression in Eq.(\ref{rfull}), is significant and rapidly changes above the threshold. The detailed behavior of the correction depends on the elastic element $S_e$ of the scattering matrix for the heavy mesons. This can be studied experimentally by comparing the yield in the threshold channel with that in a non-threshold inelastic channel. A result of such study can provide an insight into the parameters of the strong interaction between heavy mesons, and can also be used for determining the parameters of threshold resonances, such as the $Z_b$ peaks.

This work  is supported, in part, by U.S. Department of Energy Grant No. DE-SC0011842.


\begin{thebibliography}{99}
\bibitem{bellez} 
  A.~Bondar {\it et al.}  [Belle Collaboration],
  Phys.\ Rev.\ Lett.\  {\bf 108}, 122001 (2012)
  [arXiv:1110.2251 [hep-ex]].
\bibitem{bgmmv}
  A.~E.~Bondar, A.~Garmash, A.~I.~Milstein, R.~Mizuk, M.~B.~Voloshin,
  Phys.\ Rev.\  {\bf D84}, 054010 (2011).
  [arXiv:1105.4473 [hep-ph]].
  
\bibitem{ov} 
  M.~B.~Voloshin, L.~B.~Okun and ,
  JETP Lett.\  {\bf 23}, 333 (1976)
  [Pisma Zh.\ Eksp.\ Teor.\ Fiz.\  {\bf 23}, 369 (1976)].
  
\bibitem{mppr} 
  L.~Maiani, F.~Piccinini, A.~D.~Polosa and V.~Riquer,
  Phys.\ Rev.\ D {\bf 71}, 014028 (2005)
  [hep-ph/0412098].
  
\bibitem{ahw} 
  A.~Ali, C.~Hambrock and W.~Wang,
  Phys.\ Rev.\ D {\bf 85}, 054011 (2012)
  [arXiv:1110.1333 [hep-ph]].

\bibitem{ampr} 
  A.~Ali, L.~Maiani, A.~D.~Polosa and V.~Riquer,
  Phys.\ Rev.\ D {\bf 91}, no. 1, 017502 (2015)
  [arXiv:1412.2049 [hep-ph]].
  
\bibitem{mv11} 
  M.~B.~Voloshin,
  Phys.\ Rev.\ D {\bf 84}, 031502 (2011)
  [arXiv:1105.5829 [hep-ph]].
  
\bibitem{bellezbb} 
  I.~Adachi {\it et al.}  [Belle Collaboration],
  arXiv:1209.6450 [hep-ex].
\bibitem{hkmmnw} 
  C.~Hanhart, Y.~S.~Kalashnikova, P.~Matuschek, R.~V.~Mizuk, A.~V.~Nefediev and Q.~Wang,
  arXiv:1507.00382 [hep-ph].

  
\bibitem{mv13} 
  M.~B.~Voloshin,
  Phys.\ Rev.\ D {\bf 87}, no. 7, 074011 (2013)
  [arXiv:1301.5068 [hep-ph]].
  
\bibitem{lv} 
  X.~Li and M.~B.~Voloshin,
  Phys.\ Rev.\ D {\bf 87}, no. 9, 094033 (2013)
  [arXiv:1303.2949 [hep-ph]].
	
\bibitem{mp} 
  T.~Mehen and J.~W.~Powell,
  Phys.\ Rev.\ D {\bf 84}, 114013 (2011)
  [arXiv:1109.3479 [hep-ph]].
	
\bibitem{ll}
L.D. Landau and E.M. Lifshits, {\it Quantum Mechanics (Non-relativistic
Theory)}, Third Edition, Pergamon, Oxford, 1977.


\bibitem{cleo} 
  A.~Anastassov {\it et al.}  [CLEO Collaboration],
  Phys.\ Rev.\ D {\bf 65}, 032003 (2002)
  [hep-ex/0108043].	
 

\bibitem{babar} 
  J.~P.~Lees {\it et al.}  [BaBar Collaboration],
  Phys.\ Rev.\ Lett.\  {\bf 111}, no. 11, 111801 (2013)
  [arXiv:1304.5657 [hep-ex]].
	
\bibitem{pdg} 
  K.~A.~Olive {\it et al.}  [Particle Data Group Collaboration],
  Chin.\ Phys.\ C {\bf 38}, 090001 (2014).
	
\bibitem{nv} 
  J.~Nieves and M.~P.~Valderrama,
  Phys.\ Rev.\ D {\bf 86}, 056004 (2012)
  [arXiv:1204.2790 [hep-ph]].
	
\bibitem{nv11} 
  J.~Nieves and M.~P.~Valderrama,
  Phys.\ Rev.\ D {\bf 84}, 056015 (2011)
  [arXiv:1106.0600 [hep-ph]].
	
\bibitem{mpv} 
  M.~P.~Valderrama,
  Phys.\ Rev.\ D {\bf 85}, 114037 (2012)
  [arXiv:1204.2400 [hep-ph]].	

\bibitem{dlorv} 
  S.~Dubynskiy, A.~Le Yaouanc, L.~Oliver, J.-C.~Raynal and M.~B.~Voloshin,
  Phys.\ Rev.\ D {\bf 75}, 113001 (2007)
  [arXiv:0704.0293 [hep-ph]].
	
\bibitem{mv12} 
  M.~B.~Voloshin,
  Phys.\ Rev.\ D {\bf 86}, 034013 (2012)
  [arXiv:1204.1945 [hep-ph]].

\end{thebibliography}
\end{document}